\documentclass[12pt]{article}

\usepackage{amsmath,amssymb}
\usepackage{graphicx}

\newcommand{\ket}[1]{| #1 \rangle}
\newcommand{\bra}[1]{\langle #1 |}
\newcommand{\hcs}[1]{#1^\dagger #1}
\newcommand{\expv}[1]{\langle #1 \rangle}

\begin{document}

\begin{center}
{\Large\bf Hermitian conjugate measurement}
\vskip .6 cm
Hiroaki Terashima$^{1}$ and Masahito Ueda$^{2,3}$
\vskip .4 cm
{\it $^1$Department of Physics, Faculty of Education, Gunma University, \\
Maebashi, Gunma 371-8510, Japan} \\
{\it $^2$Department of Physics, University of Tokyo, \\
Bunkyo-ku, Tokyo 113-0033, Japan} \\
{\it $^3$ERATO Macroscopic Quantum Control Project, JST, \\
Bunkyo-ku, Tokyo 113-8656, Japan}
\vskip .6 cm
\end{center}

\begin{abstract}
We propose a new class of probabilistic reversing operations
on the state of a system that was disturbed by a weak measurement.
It can approximately recover the original state from
the disturbed state especially with an additional information gain
using the Hermitian conjugate of the measurement operator.
We illustrate the general scheme
by considering a quantum measurement
consisting of spin systems with an experimentally
feasible interaction and show that
the reversing operation simultaneously
increases both the fidelity to the original state
and the information gain with such a high probability of success
that their average values increase simultaneously.
\end{abstract}

\begin{flushleft}
{\footnotesize
}
\end{flushleft}

\section{Introduction}
Quantum measurement not only provides
information about a physical system
but also changes the state of the system
because of its back-action.
Although such a change in state was widely believed
to be intrinsically irreversible~\cite{LanLif77},
it has been shown that quantum measurement
is not necessarily irreversible~\cite{UedKit92},
because a certain class of measurements preserves
all the information about the system
during the measurement process.
In recent work~\cite{UedKit92,Imamog93,Royer94,UeImNa96,%
MabZol96,NieCav97,Ueda97,KoaUed99,Ban01,%
TerUed03,TerUed05,KorJor06,TerUed07}
on reversibility in quantum measurements, 
probabilistic reversing operations based on
the inverse operator of $\hat{M}$~\cite{Royer94,UeImNa96,%
Ueda97,KoaUed99,Ban01,TerUed03,TerUed05,KorJor06}
have been discussed,
where $\hat{M}$ is an operator describing
the state change due to the measurement.
That is, a second measurement is performed on the system
so that it applies $\hat{M}^{-1}$ to the system state
to cancel the effect of $\hat{M}$,
when a preferred outcome is obtained.
However, if the premeasurement state is completely recovered
using $\hat{M}^{-1}$,
the information obtained by the first measurement
is completely erased or neutralized by
the information gain from the reversing operation
(see Erratum of Ref.~\cite{Royer94}).
Recently, this type of reversing operation
has been experimentally demonstrated
using a superconducting phase qubit~\cite{KNABHL08}.

In this paper, we consider a probabilistic reversing operation
that can accomplish both approximate recovery of
the premeasurement state and additional information gain.
The operation is carried out with
the Hermitian conjugate operator of $\hat{M}$
rather than $\hat{M}^{-1}$.
Note that $\hat{M}^\dagger$ and $\hat{M}^{-1}$ are different
because $\hat{M}$ is not unitary.
However, the difference can be small if
the interaction between the system and
the measuring apparatus is sufficiently weak.
In this case, $\hat{M}^\dagger$ could approximately
cancel the state change caused by the measurement.
Moreover, a reversing operation using
$\hat{M}^\dagger$ has an advantage over
that using $\hat{M}^{-1}$ with respect to information gain.
On observing the recovery by $\hat{M}^{-1}$,
one might think that
if the premeasurement state is approximately recovered,
most of the information obtained is lost
during the reversing operation.
However, we show that
if it is approximately recovered using $\hat{M}^\dagger$,
the reversing operation \emph{increases}
rather than decreases information gain.

The additional information gain can be understood by
polar decomposition of $\hat{M}$, i.e., $\hat{M}=\hat{U}\hat{N}$,
where $\hat{U}$ is a unitary operator and
$\hat{N}$ is a nonunitary positive operator.
As shown below, 
$\hat{N}$ carries information about the system,
while $\hat{U}$ does not.
The reversing operation by $\hat{M}^\dagger$
can thus increase information gain, since
$\hat{M}^\dagger$ cancels the unitary part $\hat{U}$
but enhances the information-carrying nonunitary part $\hat{N}$ as
$\hat{M}^\dagger\hat{M}=\hat{N}^2$.
This is in contrast with the reversing operation by $\hat{M}^{-1}$,
where $\hat{M}^{-1}$ cancels not only $\hat{U}$ but also $\hat{N}$
as $\hat{M}^{-1}\hat{M}=\hat{I}$.
Of course, the premeasurement state cannot perfectly
be recovered by $\hat{M}^\dagger$,
since $\hat{N}$ disturbs the state of the system.
Nevertheless, the premeasurement state  can approximately
be recovered by $\hat{M}^\dagger$
as long as the state disturbance by $\hat{N}$ is
much smaller than that by $\hat{U}$.
We shall show such a physical example using
spin systems with Ising-type interaction.

An approximate recovery with additional information gain
was first discussed in Ref.~\cite{TerUed05}.
However, the paper did not identify the reason for the information gain
because it focused on a reversing operation by $\hat{M}^{-1}$.
Similarly, an approximate recovery with purity gain
(instead of information gain)
was discussed in Ref.~\cite{TerUed07}
for a system weakly interacting with the environment
by regarding the interaction with the environment
as a measurement.
However, the reversing operation in that case requires
the average over the outcome of the ``measurement,''
since the environment does not refer to the outcome.
This obscures the nature of the operator that contributes to the purity gain.
Therefore, here we clarify
the reason for the information gain,
together with the property of the operator that is required to
achieve the information gain.

This paper is organized as follows:
Section~\ref{sec:measure} describes
the general formulation of quantum measurement
and introduces fidelity loss and information gain
due to measurement.
Section~\ref{sec:conj} defines a Hermitian conjugate measurement
together with the reversing measurement scheme.
Section~\ref{sec:weak} shows that
in the case of weak measurement,
the Hermitian conjugate measurement
achieves both approximate recovery of
the premeasurement state and additional information gain.
Section~\ref{sec:example} considers
a quantum measurement of a spin-$s$ system
using a spin-$j$ probe as an example.
Section~\ref{sec:conclude} summarizes our results.

\section{\label{sec:measure}Quantum Measurement,
Fidelity, and Information Gain}
A quantum measurement is generally
described~\cite{DavLew70,NieChu00} by a set of
linear operators $\{\hat{M}_m\}$, called measurement operators,
that satisfy the completeness condition
\begin{equation}
\sum_m\hcs{\hat{M}_m}=\hat{I},
\label{eq:condm}
\end{equation}
where $\hat{I}$ is the identity operator.
If the system to be measured is in a state $\hat{\rho}$,
the measurement yields outcome $m$ with probability
\begin{equation}
 p_m=\mathrm{Tr}(\hat{\rho}\,\hat{M}_m^\dagger\hat{M}_m),
\end{equation}
and for each outcome $m$ the state of the system is changed into
\begin{equation}
\hat{\rho}_m=\frac{1}{p_m}
   \hat{M}_m \,\hat{\rho}\hat{M}_m^\dagger.
\label{eq:defpsim}
\end{equation}
We can always construct a quantum measurement
described by a given set of operators $\{\hat{M}_m\}$,
using a measuring apparatus whose initial state, interaction,
and observable are appropriately chosen~\cite{NieChu00}.

Provided that the dimension of the support is finite,
any linear operator $\hat{M}_m$ can uniquely be
decomposed by \emph{left} polar decomposition into
\begin{equation}
  \hat{M}_m=\hat{U}_m \hat{N}_m,
\label{eq:decomp}
\end{equation}
where $\hat{U}_m$ is a unitary operator and
$\hat{N}_m\equiv\sqrt{\hat{M}_m^\dagger\hat{M}_m}$ is
a positive operator.
The operators $\{\hat{N}_m\}$ also describe a quantum measurement
because they are linear and
satisfy $\sum_m\hcs{\hat{N}_m}=\hat{I}$.
The measurement described by $\{\hat{N}_m\}$ gives
the same amount of information gain
as the measurement $\{\hat{M}_m\}$
but changes the state as little as possible.
This is because
the probability $p_m=\mathrm{Tr}(\hat{\rho}\hat{N}_m^2)$
does not depend on $\hat{U}_m$.
The unitary part, $\hat{U}_m$, is thus irrelevant to
the information gain\ and contributes only to the state change.
Unfortunately,
we cannot always perform this optimal measurement $\{\hat{N}_m\}$
since available interactions between the system and
the measuring apparatus are subject to experimental constraints.

In making the polar decomposition (\ref{eq:decomp})
of the measurement operator,
we have assumed that
the system's Hilbert space is finite-dimensional,
because a linear operator on an infinite-dimensional Hilbert space
cannot always be decomposed
by polar decomposition~\cite{WuZha00}.
This assumption is not particularly restrictive,
owing to the existence of a physical cutoff.
For example, in photon counting~\cite{UeImOg90},
the measurement process that detects one photon
with a photodetector is described by
the annihilation operator, $\hat{a}$, of the photon;
however, it has been shown that
such an annihilation operator does not have
polar decomposition~\cite{Fujika95}.
Note that the Hilbert space of the photon field is
infinite-dimensional, since it is spanned by the eigenstates $\ket{n}$
of the photon-number operator $\hat{a}^\dagger\hat{a}$
with $n=0,1,2,\ldots$.
Even in this case,
an effective upper bound on the photon number $n_{\max}$ can be
introduced by considering an actual experimental setup.
Truncating the Hilbert space $\{\ket{n}\}$
to finite dimensions $n=0,1,2,\ldots,n_{\max}$,
we can consider an approximate polar decomposition
as in Eq.~(\ref{eq:decomp}).

To evaluate the amount of information
obtained by a \emph{single} measurement outcome,
suppose that the premeasurement state $\hat{\rho}$ is known to
be one of the predefined states $\{\hat{\rho}(a)\}$
with equal probability, $p(a)=1/N$, where $a=1,\ldots,N$.
Since the premeasurement state is usually an arbitrary unknown state
in quantum measurement,
$\{\hat{\rho}(a)\}$ is essentially an infinite set ($N\to\infty$).
This contrasts with the case of
quantum state discrimination~\cite{Peres88,DuaGuo98},
in which $N$ cannot be greater than the dimension of the
Hilbert space due to the linear independence of $\{\hat{\rho}(a)\}$.
The Shannon entropy associated with the system is
initially
\begin{equation}
  H_0=-\sum_a p(a)\log_2 p(a)=\log_2 N,
\end{equation}
which is a measure of the lack of information about the system.

The measurement $\{\hat{M}_m\}$ is then performed
to obtain information about the system.
If the premeasurement state is $\hat{\rho}(a)$,
the measurement yields an outcome $m$ with probability
\begin{equation}
 p(m|a)
  = \expv{\hat{M}_m^\dagger\hat{M}_m}_a
  = \expv{\hat{N}_m^2}_a,
\end{equation}
where the bracket with subscript $a$ denotes
\begin{equation}
  \expv{\hat{O}}_a\equiv
     \mathrm{Tr}\left[\hat{\rho}(a)\hat{O}\right].
\end{equation}
The total probability for outcome $m$ is thus
\begin{equation}
  p(m)=\sum_a  p(m|a)p(a)
      =\frac{1}{N}\sum_a\expv{\hat{N}_m^2}_a
      =\overline{\expv{\hat{N}_m^2}},
\label{eq:defpm}
\end{equation}
where the overline denotes the average over $a$,
\begin{equation}
   \overline{f} \equiv \frac{1}{N}\sum_a f(a).
\label{eq:overline}
\end{equation}
Conversely, given outcome $m$,
we can find the probability that
the premeasurement state is $\hat{\rho}(a)$ by
\begin{equation}
 p(a|m)=\frac{p(m|a)p(a)}{p(m)}
\label{eq:bayes}
\end{equation}
from Bayes' rule.
This indicates that
the Shannon entropy after measurement
with outcome $m$ is
\begin{equation}
  H(m) =-\sum_a p(a|m)\log_2 p(a|m).
\end{equation}
Therefore,
the amount of information obtained from
outcome $m$ is evaluated by
\begin{equation}
I(m)=H_0-H(m)=
  \frac{\overline{\expv{\hat{N}_m^2}\log_2 \expv{\hat{N}_m^2}}-
        \overline{\expv{\hat{N}_m^2}}\log_2\overline{\expv{\hat{N}_m^2}}}
  {\overline{\expv{\hat{N}_m^2}}},
\label{eq:defim}
\end{equation}
owing to the assumption that $p(a)=1/N$ does not depend on $a$.
The mean information gain after the measurement is given by
\begin{equation}
  I=\sum_m p(m) I(m).
\label{eq:imbar}
\end{equation}

On the other hand,
the state change caused by the measurement
can be evaluated in terms of the fidelity~\cite{Uhlman76,NieChu00}
between the premeasurement and postmeasurement states.
If the premeasurement state is $\hat{\rho}(a)$ and
the measurement outcome is $m$,
the postmeasurement state is given by
\begin{equation}
\hat{\rho}(m,a)=\frac{1}{p(m|a)}
    \hat{M}_m \,\hat{\rho}(a)\hat{M}_m^\dagger.
\label{eq:rhoma}
\end{equation}
The fidelity between the premeasurement and
postmeasurement states then becomes
\begin{equation}
 F(m,a) =\mathrm{Tr}
   \sqrt{\sqrt{\hat{\rho}(a)}\;\hat{\rho}(m,a)\,\sqrt{\hat{\rho}(a)}},
\end{equation}
with $0\le F(m,a)\le1$.
The more drastically the measurement changes the state of the system,
the smaller the fidelity becomes.
Since $a$ is unknown to us,
the fidelity after the measurement with outcome $m$ is
evaluated using the probability in Eq.~(\ref{eq:bayes}) by
\begin{equation}
 F(m)=\sum_a p(a|m) F(m,a).
\label{eq:deffm}
\end{equation}
The mean fidelity after measurement is given by
\begin{equation}
  F=\sum_m p(m) F(m).
\label{eq:fmbar}
\end{equation}

\section{\label{sec:conj}Hermitian Conjugate Measurement}
To undo the state change caused by measurement,
a reversing measurement scheme was proposed in Ref.~\cite{UeImNa96} based on
the inverse of the measurement operator.
In this scheme, depending on the outcome $m$ of the measurement,
another measurement, called a reversing measurement,
is performed on the postmeasurement state (\ref{eq:defpsim}) of the system.
The reversing measurement is described by a set of
measurement operators $\{\hat{R}^{(m)}_\nu\}$
that satisfy~\cite{UeImNa96}
\begin{equation}
\sum_\nu\hat{R}^{(m)\dagger}_\nu\hat{R}^{(m)}_\nu=\hat{I}
\label{eq:revunit}
\end{equation}
and
\begin{equation}
 \hat{R}^{(m)}_{\nu_0}=\lambda_m\, \hat{M}_m^{-1},
 \qquad 0<|\lambda_m|^2\le \inf_{\hat{\rho}}\, p_m
\end{equation}
for a particular $\nu_0$,
where $\nu$ denotes the outcome of the reversing measurement
and $\lambda_m$ is a complex number.
The upper bound for $\lambda_m$ is
determined by the condition (\ref{eq:revunit}), namely,
$\expv{\hat{R}^{(m)\dagger}_{\nu_0}\hat{R}^{(m)}_{\nu_0}}\le 1$
for any $\hat{\rho}$~\cite{KoaUed99}.
Thus, the reversing measurement restores the premeasurement state
if the measurement outcome is $\nu_0$.

In our situation with the predefined states $\{\hat{\rho}(a)\}$,
when an outcome $\nu$ is obtained from the reversing measurement
on the state (\ref{eq:rhoma}),
the state of the system becomes
\begin{equation}
\hat{\rho}(m,\nu,a)=\frac{1}{p(m,\nu|a)}\,
   \hat{R}^{(m)}_\nu\hat{M}_m\,\hat{\rho}(a) \,
   \hat{M}_m^\dagger \hat{R}^{(m)\dagger}_\nu,
\end{equation}
where
\begin{equation}
 p(m,\nu|a)\equiv 
    \expv{\hat{M}_m^\dagger\hat{R}^{(m)\dagger}_\nu
          \hat{R}^{(m)}_\nu\hat{M}_m}_a
\end{equation}
is the joint probability for obtaining
the set of outcomes $(m,\nu)$ for the two successive measurements
$\{\hat{M}_m\}$ and $\{\hat{R}^{(m)}_\nu\}$.
Conversely, given outcomes $(m,\nu)$,
we can find the probability that
the premeasurement state is $\hat{\rho}(a)$, with
\begin{equation}
 p(a|m,\nu)=\frac{p(m,\nu|a)p(a)}{p(m,\nu)},
\label{eq:bayes2}
\end{equation}
where $p(m,\nu)$ is the total probability
for the set of outcomes $(m,\nu)$:
\begin{equation}
  p(m,\nu)=\sum_a  p(m,\nu|a)p(a).
\end{equation}
The information gain then becomes
\begin{equation}
  I(m,\nu)=H_0-H(m,\nu),
\label{eq:imnu}
\end{equation}
with $H(m,\nu)$ being the Shannon entropy after
the reversing measurement:
\begin{equation}
  H(m,\nu) =-\sum_a p(a|m,\nu)\log_2 p(a|m,\nu).
\end{equation}
On the other hand, the fidelity after
the reversing measurement is expressed as
\begin{equation}
 F(m,\nu)=\sum_a p(a|m,\nu) F(m,\nu,a),
\label{eq:fmnu}
\end{equation}
where $p(a|m,\nu)$ is given in Eq.~(\ref{eq:bayes2}) and
$F(m,\nu,a)$ is the fidelity defined by
\begin{equation}
F(m,\nu,a) \equiv\mathrm{Tr}
 \sqrt{\sqrt{\hat{\rho}(a)}\;\hat{\rho}(m,\nu,a)\,\sqrt{\hat{\rho}(a)}}.
\end{equation}

If outcome $\nu$ is that $\nu_0$ for which
the premeasurement state is recovered,
fidelity (\ref{eq:fmnu}) and information gain (\ref{eq:imnu})
reduce to
\begin{align}
F(m,\nu_0) &=1, \label{eq:perfrevf} \\
I(m,\nu_0) &=0, \label{eq:perfrevi} 
\end{align}
since $\hat{R}^{(m)}_{\nu_0}$ is proportional to
the inverse operator of $\hat{M}_m$,
\begin{equation}
\hat{R}^{(m)}_{\nu_0}\hat{M}_m \propto \hat{I}.
\label{eq:invrel}
\end{equation}
That is,
if the particular outcome $\nu_0$ is obtained
by the reversing measurement,
the unknown original state $\hat{\rho}(a)$ is
perfectly recovered because
the inverse operator of $\hat{M}_m$ is applied to
the system's state.
However, when perfect recovery is achieved,
the information obtained by the first measurement is completely
lost by the reversing measurement, $p(a|m,\nu_0)=p(a)$,
because the information concerning
the premeasurement state is not reflected
in the joint probability distribution
for the perfect recovery~\cite{UeImNa96};
i.e., $p(m,\nu_0|a)=|\lambda_m|^2$ does not
depend on $\hat{\rho}(a)$.

Now, we consider a reversing operation that is
based on the Hermitian conjugate of the measurement operator.
That is, instead of
the reversing measurement $\{\hat{R}^{(m)}_\nu\}$,
we perform a measurement $\{\hat{C}^{(m)}_\mu\}$ satisfying
\begin{equation}
\sum_\mu\hat{C}^{(m)\dagger}_\mu\hat{C}^{(m)}_\mu=\hat{I}
\label{eq:condc}
\end{equation}
and
\begin{equation}
 \hat{C}^{(m)}_{\mu_0}=\kappa_m\,\hat{M}_m^\dagger,
 \qquad 0<|\kappa_m|^2\le \left(\sup_{\hat{\rho}}\, p_m\right)^{-1}
\label{eq:defcmu}
\end{equation}
with a complex number $\kappa_m$ for a particular outcome $\mu_0$.
The upper bound for $\kappa_m$ is
determined by the condition
$\expv{\hat{C}^{(m)\dagger}_{\mu_0}\hat{C}^{(m)}_{\mu_0}}\le 1$
for any $\hat{\rho}$, which is equivalent to the condition
$\expv{\hat{C}^{(m)}_{\mu_0}\hat{C}^{(m)\dagger}_{\mu_0}}\le 1$
for any $\hat{\rho}$ because of polar decomposition (\ref{eq:decomp}).
We shall refer to $\{\hat{C}^{(m)}_\mu\}$ as
a Hermitian conjugate measurement.

In our situation with $\{\hat{\rho}(a)\}$,
when outcome $\mu$ is obtained by the Hermitian conjugate measurement
on state (\ref{eq:rhoma}),
the state of the system becomes
\begin{equation}
\hat{\rho}(m,\mu,a)=\frac{1}{p(m,\mu|a)}\,
    \hat{C}^{(m)}_\mu\hat{M}_m\,\hat{\rho}(a)\,
    \hat{M}_m^\dagger \hat{C}^{(m)\dagger}_\mu,
\end{equation}
where
\begin{equation}
 p(m,\mu|a)\equiv
   \expv{\hat{M}_m^\dagger\hat{C}^{(m)\dagger}_\mu
         \hat{C}^{(m)}_\mu\hat{M}_m}_a
\end{equation}
is the joint probability for
the set of outcomes $(m,\mu)$.
We define fidelity $F(m,\mu)$ and information gain $I(m,\mu)$
as in the case of reversing measurement,
replacing $\hat{R}^{(m)}_\nu$ with $\hat{C}^{(m)}_\mu$.
If the outcome $\mu$ is the preferred one $\mu_0$,
the fidelity and information gain reduce to
\begin{align}
 F(m,\mu_0) &=\frac{1}{\overline{\expv{\hat{N}_m^4}}}
    \overline{\sqrt{\expv{\hat{N}_m^4}}\,\expv{\hat{N}_m^2}},
  \label{eq:fmmu0} \\
 I(m,\mu_0) &=
\frac{\overline{\expv{\hat{N}_m^4}\log_2 \expv{\hat{N}_m^4}}-
      \overline{\expv{\hat{N}_m^4}}\log_2\overline{\expv{\hat{N}_m^4}}}
{\overline{\expv{\hat{N}_m^4}}},
  \label{eq:immu0}
\end{align}
since from Eqs.~(\ref{eq:defcmu}) and (\ref{eq:decomp}) we have
\begin{align}
\hat{C}^{(m)}_{\mu_0}\hat{M}_m \propto \hat{N}_m^2.
\label{eq:conjrel}
\end{align}
In the next section, we show that
if the preferred outcome $\mu_0$ is obtained
by the Hermitian conjugate measurement,
the unknown original state $\hat{\rho}(a)$ is approximately recovered
with additional information gain for a weak measurement.

\section{\label{sec:weak}Simultaneous State Recovery
and Information Gain for a Weak Measurement}
We consider the case of a measurement $\{\hat{M}_m\}$
that provides only a small amount of information,
e.g., measurement by an apparatus having a weak interaction
with the system.
In this case,
$\hat{N}_m$ in Eq.~(\ref{eq:decomp}) can be expressed as
\begin{equation}
  \hat{N}_m\equiv
    q_m\left(\hat{I}+\hat{\epsilon}_m\right),
\label{eq:weakn}
\end{equation}
where $q_m$ is a positive number
and $\hat{\epsilon}_m$ is a small Hermitian operator.
It follows from Eq.~(\ref{eq:condm}) that
$\{q_m\}$ and $\{\hat{\epsilon}_m \}$ satisfy
\begin{align}
 & \sum_m q_m^2 =1, \\
 & \sum_m q_m^2
\left(2\hat{\epsilon}_m+ \hat{\epsilon}_m^2\right)=0.
\end{align}
Then, up to the order of $\hat{\epsilon}_m^2$,
the information gain in Eq.~(\ref{eq:defim}) and
its mean in Eq.~(\ref{eq:imbar}) are calculated to be
\begin{align}
& I(m) \simeq 2 V_I\left(\hat{\epsilon}_m\right), \label{eq:weakim} \\
& I \simeq 2\sum_m q_m^2  V_I\left(\hat{\epsilon}_m\right),
\label{eq:weakibar}
\end{align}
where $V_I\left(\hat{\epsilon}_m\right)$ is
a variance defined by
\begin{equation}
  V_I\left(\hat{\epsilon}_m\right)
   \equiv \overline{\expv{\hat{\epsilon}_m}^2}-
        \left(\overline{\expv{\hat{\epsilon}_m}}\right)^2
   =\overline{\left(\expv{\hat{\epsilon}_m}-
        \overline{\expv{\hat{\epsilon}_m}}\right)^2} \ge 0.
\end{equation}
This is a classical variance with respect to $a$
of the quantum average $\expv{\hat{\epsilon}_m}_a$.

On the other hand, a weak measurement does not
necessarily imply a small change in the system state,
since the state change depends
not only on $\hat{N}_m$ but also on $\hat{U}_m$
in Eq.~(\ref{eq:decomp}).
In general, $\hat{U}_m$ can be written as
\begin{equation}
  \hat{U}_m \equiv e^{i\gamma_m}e^{i\hat{\Gamma}_m},
\label{eq:defgamma}
\end{equation}
where $\gamma_m$ is a real number and
$\hat{\Gamma}_m$ is a Hermitian operator.
Note that, even if the interaction
between the system and the measuring apparatus is weak,
$\hat{\Gamma}_m$ can be large if
the degrees of freedom of
the system or those of the measuring apparatus
are large~\cite{TerUed07}, as shown below.
When all $\hat{\rho}(a)$'s are pure,
$\hat{\rho}(a)=\ket{\psi(a)}\bra{\psi(a)}$,
we obtain the fidelity from Eq.~(\ref{eq:deffm}) and
its mean from Eq.~(\ref{eq:fmbar}) as
\begin{align}
& F(m) \simeq 
    \overline{\left|\bra{\psi}e^{i\hat{\Gamma}_m}\ket{\psi}\right|}
   \left[\,1+O(\hat{\epsilon}_m)\, \right], \label{eq:fm1st} \\
& F \simeq 
  \sum_m q_m^2 \overline{\left|\bra{\psi}e^{i\hat{\Gamma}_m}\ket{\psi}\right|}
  \left[\,1+O(\hat{\epsilon}_m)\, \right]. \label{eq:fbar1st}
\end{align}
Equations (\ref{eq:fm1st}) and (\ref{eq:fbar1st}) show that
the fidelity can almost vanish if $\hat{\Gamma}_m$ is large enough,
even though large $\hat{\Gamma}_m$ does not always imply small $F(m)$.
Below, we consider a measurement
that provides a small amount of information
through Eq.~(\ref{eq:weakn}), despite the fact that
it drastically changes the state of the system, such that
\begin{equation}
  \frac{1-F(m)}{1-F_{\mathrm{opt}}(m)} > 4,
\label{eq:assumpfm}
\end{equation}
where $F_{\mathrm{opt}}(m)$ would be the fidelity
if the measurement were optimal, i.e., $\hat{\Gamma}_m=0$.
The explicit form of $F_{\mathrm{opt}}(m)$ is
\begin{equation}
  F_{\mathrm{opt}}(m) =\frac{1}{\overline{\expv{\hat{N}_m^2}}}
    \overline{\sqrt{\expv{\hat{N}_m^2}}\,\expv{\hat{N}_m}}
  \simeq 1-\frac{1}{2}V_F\left(\hat{\epsilon}_m\right),
\label{eq:foptm}
\end{equation}
with $V_F\left(\hat{\epsilon}_m\right)$ being
a variance defined by
\begin{equation}
 V_F\left(\hat{\epsilon}_m\right)
  \equiv \overline{\expv{\hat{\epsilon}_m^2}}-
     \overline{\expv{\hat{\epsilon}_m}^2}
 =\overline{\expv{\left(\hat{\epsilon}_m-
   \expv{\hat{\epsilon}_m}\right)^2}}
  \ge 0.
\end{equation}
This is a classical average over $a$
of the quantum variance
$\expv{\left(\hat{\epsilon}_m-\expv{\hat{\epsilon}_m}_a\right)^2}_a$.

From Eqs.~(\ref{eq:fmmu0}) and (\ref{eq:immu0}),
the fidelity and information gain after the Hermitian conjugate measurement
with the preferred outcome $\mu_0$ can be calculated
up to the order of $\hat{\epsilon}_m^2$ to be
\begin{align}
 F(m,\mu_0) &\simeq 1-2 V_F\left(\hat{\epsilon}_m\right), \\
 I(m,\mu_0) &\simeq 8 V_I\left(\hat{\epsilon}_m\right).
 \label{eq:recoveri}
\end{align}
Note that as long as higher-order terms can be ignored,
\begin{equation}
   F(m,\mu_0) > F(m)
\end{equation}
by the assumption made in Eq.~(\ref{eq:assumpfm}).
This means that
the Hermitian conjugate measurement approximately recovers
the original state $\hat{\rho}(a)$.
Moreover, it follows from Eqs.~(\ref{eq:weakim}) and (\ref{eq:recoveri}) that
the Hermitian conjugate measurement \emph{simultaneously}
enhances the information gain by a factor of four, since
\begin{equation}
  I(m,\mu_0) \simeq 4 I(m).
\end{equation}
Such an approximate recovery occurs because
$\hat{U}_m^\dagger$ in $\hat{C}^{(m)}_{\mu_0}$
cancels the large disturbance caused
by the unitary part $\hat{U}_m$ in $\hat{M}_m$,
while the additional information gain is obtained because
the composition of $\hat{M}_m$ and $\hat{C}^{(m)}_{\mu_0}$
results in the optimal measurement $\hat{N}_m$ being applied twice,
as shown in Eq.~(\ref{eq:conjrel}).
The state recovery of Hermitian conjugate measurement
presents a sharp contrast to
that of the reversing measurement shown in
Eqs.~(\ref{eq:perfrevf}) and (\ref{eq:perfrevi}),
in which the reversing measurement perfectly recovers
the original state $\hat{\rho}(a)$,
but completely obliterates the information $I(m)$.
The recovery with information loss
occurs because $\hat{R}^{(m)}_{\nu_0}$
contains not only $\hat{U}_m^\dagger$, which cancels $\hat{U}_m$,
but also $\hat{N}_m^{-1}$, which cancels
the nonunitary part $\hat{N}_m$ in $\hat{M}_m$,
as in Eq.~(\ref{eq:invrel}).

One might think that the probability for
an approximate recovery is very low, and
if an average over the outcome $\mu$ is taken,
the fidelity increases with a decrease in information gain.
However, the preferred outcome $\mu_0$
is more probable when 
the outcome $m$ of the measurement $\{\hat{M}_m\}$
occurs with high probability.
In fact, given outcome $m$,
the conditional probability for outcome $\mu$ of
the Hermitian conjugate measurement $\{\hat{C}^{(m)}_\mu\}$ is given by
$p(\mu|m)=p(m,\mu)/p(m)$,
which, for the preferred outcome $\mu_0$, reduces to
\begin{equation}
 p(\mu_0|m)\simeq 
   |\kappa_m|^2\left\{\,p(m)+4q_m^2\left[
    V_F\left(\hat{\epsilon}_m\right)+V_I\left(\hat{\epsilon}_m\right)
   \right]\right\}.
\label{eq:condprob}
\end{equation}
This indicates that, when $p(m)$ is large,
$p(\mu_0|m)$ is also large.
Discussing the mean fidelity and information gain
conditioned by outcome $m$,
\begin{align}
 F'(m) &\equiv\sum_\mu p(\mu|m)\,F(m,\mu),
  \label{eq:deffmp} \\
 I'(m) &\equiv\sum_\mu p(\mu|m)\,I(m,\mu),
  \label{eq:defimp}
\end{align}
we must specify $\hat{C}^{(m)}_\mu$'s other than $\mu=\mu_0$.
Here, we consider a minimal model,
where the only two possible outcomes of the Hermitian conjugate measurement
are $\mu=\mu_0$ and $\mu=\mu_1$.
Then, the measurement operator for $\mu=\mu_1$ is chosen as
\begin{equation}
\hat{C}_{\mu_1}^{(m)}=\sqrt{1-a_m^2}
  \left(\hat{I}-\frac{a_m^2}{1-a_m^2} 
   \hat{\epsilon}_m -\frac{a_m^2}{2(1-a_m^2)^2} 
   \hat{\epsilon}_m^2
  \right)\,\hat{U}_m^\dagger,
\end{equation}
where $a_m^2\equiv |\kappa_m|^2q_m^2$, and we assume that
$a_m^2\hat{\epsilon}_m/(1-a_m^2)$ is small,
so that condition (\ref{eq:condc}) is satisfied
up to the order of $\hat{\epsilon}_m^2$.
When the outcome of the Hermitian conjugate measurement
$\{\hat{C}^{(m)}_\mu\}$ is $\mu_1$,
the fidelity and the information gain become
\begin{align}
 F(m,\mu_1) &\simeq 1-\frac{1}{2}
    \left(1-\frac{a_m^2}{1-a_m^2}\right)^2\,
     V_F\left(\hat{\epsilon}_m\right), \\
 I(m,\mu_1) &\simeq 2\left(1-\frac{a_m^2}{1-a_m^2}\right)^2\,
      V_I\left(\hat{\epsilon}_m\right).
\end{align}
In this case, the Hermitian conjugate measurement
decreases the information gain $I(m,\mu_1)<I(m)$
from Eq.~(\ref{eq:weakim}).
The mean fidelity (\ref{eq:deffmp}) and
information gain (\ref{eq:defimp}) 
after the Hermitian conjugate measurement are then given by
\begin{align}
 F'(m) &\simeq 1-\frac{1}{2(1-a_m^2)}\,V_F\left(\hat{\epsilon}_m\right), \\
 I'(m) &\simeq \frac{2}{1-a_m^2}\,V_I\left(\hat{\epsilon}_m\right),
\end{align}
which imply $I'(m)> I(m)$ and $F'(m) > F(m)$ if $a_m^2<3/4$
from Eq.~(\ref{eq:assumpfm}).
Therefore, the Hermitian conjugate measurement, on average,
increases both the fidelity and information gain.
We can obtain the same conclusion even after
the averages over $m$ are taken:
\begin{align}
F' &\equiv \sum_m p(m) F'(m) > F,
\label{eq:deffbarp} \\
I' &\equiv \sum_m p(m) I'(m)> I.
\label{eq:defibarp}
\end{align}

\section{\label{sec:example}Example: Ising-type Interaction}
As an example,
we consider a quantum measurement on a spin-$s$ system
described by spin operators
$\{\hat{S}_{x},\hat{S}_{y},\hat{S}_{z}\}$.
We assume that we have no \textit{a priori}
information about the state of the system
except that it is a pure state.
This means that the set of predefined states,
$\{\hat{\rho}(a)\}$, consists of all possible pure states.
That is, $\hat{\rho}(a)$ can be written as
$\hat{\rho}(a)=\ket{\psi(a)}\bra{\psi(a)}$ by a state vector
\begin{equation}
\ket{\psi(a)}=\sum_{\sigma} c_\sigma(a) \ket{\sigma},
\label{eq:pure}
\end{equation}
where $\ket{\sigma}$ is the eigenstate of $\hat{S}_z$
with eigenvalue $\sigma$ ($=-s,-s+1,\ldots,s-1,s$)
and $c_\sigma(a)$'s obey
the normalization condition $\sum_\sigma |c_\sigma(a)|^2=1$.

To obtain information about the system's state,
we perform a measurement using a spin-$j$ probe (measuring apparatus)
described by spin operators
$\{\hat{J}_{x},\hat{J}_{y},\hat{J}_{z}\}$.
The measurement proceeds as follows.
The probe is first prepared in a coherent spin state
$\ket{\theta,\pi/2}$~\cite{ACGT72}, which is
the eigenstate of the spin component
$\hat{J}_{y}\sin\theta+\hat{J}_{z}\cos\theta$
with eigenvalue $j$.
The probe then interacts with the system
via an interaction Hamiltonian
\begin{equation}
   H_\mathrm{int}=\alpha \hat{J}_{z}\hat{S}_{z},
\label{eq:int}
\end{equation}
where $\alpha$ is a real constant.
This $\hat{J}_{z}\hat{S}_{z}$-type
interaction has direct relevance to the experimental situations
in Refs.~\cite{HapMat67,KuBiMa98,THTTIY99,KMJYEB99,TITKYT05}.
After interaction during time $t$, a unitary operator
\begin{equation}
\hat{U}_\mathrm{p}=e^{-i\pi\hat{J}_{y}/2}
\end{equation}
is applied to the probe.
Finally, we obtain outcome $m$ ($=-j,-j+1,\ldots,j-1,j$)
by performing a projective measurement
on the probe observable $\hat{J}_{z}$.
The outcome $m$ then provides some information
about the state $\hat{\rho}(a)$.
The measurement process is described
by the set of measurement operators~\cite{TerUed05}
\begin{equation}
\hat{M}_m=\hat{T}_m(\theta)
         \equiv\sum_\sigma a_{m\sigma}^{(j)}(\theta)\,
\ket{\sigma}\bra{\sigma},
\end{equation}
where
\begin{align}
 a_{m\sigma}^{(j)}(\theta) &= \frac{e^{-ij\pi/2}}{2^j}
      \sqrt{\frac{(2j)!}{(j+m)!(j-m)!}}  \notag \\
  &  \qquad{}\times
     \left( e^{-ig\sigma}\cos\frac{\theta}{2}
     +ie^{ig\sigma}\sin\frac{\theta}{2} \right)^{j-m} \notag \\
  &  \qquad{}\times
     \left( e^{-ig\sigma}\cos\frac{\theta}{2}
     -ie^{ig\sigma}\sin\frac{\theta}{2} \right)^{j+m}
\end{align}
with $g\equiv\alpha t/2$ being the effective strength of the interaction.
When the interaction is weak, $\hat{N}_m$
in the decomposition of $\hat{M}_m$ in Eq.~(\ref{eq:decomp})
can be written as in Eq.~(\ref{eq:weakn}), with
\begin{align}
 q_m  &= \frac{1}{2^j}
      \sqrt{\frac{(2j)!}{(j+m)!(j-m)!}}, \\
 \hat{\epsilon}_m &\simeq 
  2gm\sin\theta\, \hat{S}_z+O(g^2), \label{eq:epsi}
\end{align}
and $\hat{U}_m$
can be written as in Eq.~(\ref{eq:defgamma}), with
\begin{align}
  \gamma_m &=-\frac{j\pi}{2}-m\theta, \\
  \hat{\Gamma}_m &\simeq
  -2gj\cos\theta\,\hat{S}_z+O(g^2). \label{eq:gammam}
\end{align}
Since the probability for outcome $m$ is $p(m)\simeq q_m^2+O(g)$
from Eq.~(\ref{eq:defpm}),
the expectation value and variance
of the outcome are given by
\begin{align}
 \bar{m} \equiv &\sum_m p(m)\, m \simeq 0+O(g), \label{eq:mbar} \\
 (\delta m)^2 \equiv &\sum_m p(m) (m-\bar{m})^2 
                      \simeq \frac{j}{2}+O(g),\label{eq:delm}
\end{align}
respectively.
Comparing Eqs.~(\ref{eq:mbar}) and (\ref{eq:delm})
with Eq.~(\ref{eq:epsi}),
we find that $\hat{\epsilon}_m \sim O(g\sqrt{j})$.
In contrast, Eq.~(\ref{eq:gammam}) shows that
$\hat{\Gamma}_m \sim O(gj)$.
Therefore, even if $\hat{\epsilon}_m$ is small,
$\hat{\Gamma}_m$ can be large for large values of $j$.
In the following discussion,
we shall consider such a situation by assuming that
$g$ is so small that
\begin{equation}
 \frac{2}{3}g^2s(s+1)j\sin^2\theta  \ll 1,
\label{eq:assump1}
\end{equation}
but $j$ is so large that
$\hat{U}_m$ differs greatly from the identity operator,
\begin{equation}
  |2gj\cos\theta|  \sim \pi.
\label{eq:assump2}
\end{equation}

Substituting Eq.~(\ref{eq:epsi}) into
Eqs.~(\ref{eq:weakim}) and (\ref{eq:weakibar}),
we obtain the information gain and its mean to the order of $g^2$ as
\begin{align}
& I(m) \simeq \frac{4}{3}g^2s\, m^2\sin^2\theta, \label{eq:exampleim} \\
& I \simeq \frac{2}{3}g^2sj\sin^2\theta,
\end{align}
where we have used
\begin{equation}
 V_I(\hat{S}_z) = \frac{1}{6}s
\label{eq:varii}
\end{equation}
(see Appendix \ref{sec:vari}).
On the other hand, we cannot expand the fidelity
in Eq.~(\ref{eq:deffm}) and its mean in Eq.~(\ref{eq:fmbar})
in terms of $g$ when $\hat{\Gamma}_m$ is large.
If we \emph{formally} expand them,
they are given by
\begin{align}
& F(m)  \simeq 1-\frac{1}{3}g^2s(2s+1)
  \left(j^2\cos^2\theta+m^2\sin^2\theta\right),  \label{eq:weakfm} \\
&  F\simeq 1-\frac{1}{3}g^2 s(2s+1)
   \left(j^2\cos^2\theta+\frac{j}{2}\sin^2\theta\right),
   \label{eq:weakf}
\end{align}
respectively, since the variance $V_F(\hat{S}_z)$ is
calculated to be
\begin{equation}
 V_F(\hat{S}_z) = \frac{1}{6}s(2s+1)
\label{eq:varif}
\end{equation}
(see Appendix \ref{sec:vari}).
Compared to Eq.~(\ref{eq:weakfm}),
the optimal fidelity (\ref{eq:foptm}) can
be expanded in terms of $g$ as
\begin{equation}
F_{\mathrm{opt}}(m)\simeq 1-\frac{1}{3}g^2s(2s+1)m^2\sin^2\theta,
\end{equation}
without the term of order $g^2j^2$ originating from $\hat{\Gamma}_m$.

\begin{figure}
\begin{center}
\includegraphics[scale=0.65]{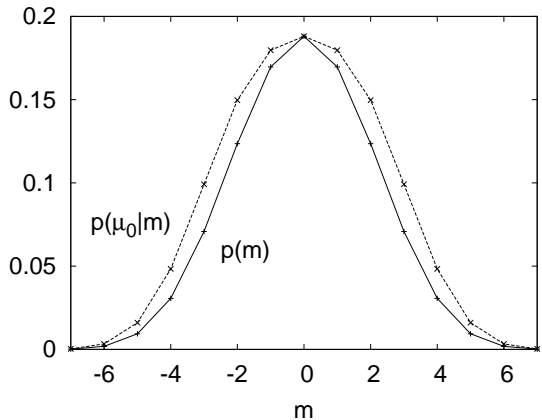}
\end{center}
\caption{\label{fig1}Probability $p(m)$ of
obtaining outcome $m$ for measurement $\{\hat{M}_m\}$
and probability $p(\mu_0|m)$ of obtaining
the preferred outcome $\mu_0=m$
for the Hermitian conjugate measurement $\{\hat{C}^{(m)}_\mu\}$
conditioned by the first outcome $m$,
with $s=1/2$, $j=7$, $g=0.25$, and $\theta=\pi/6$.
}
\end{figure}

\begin{figure}
\begin{center}
\includegraphics[scale=0.65]{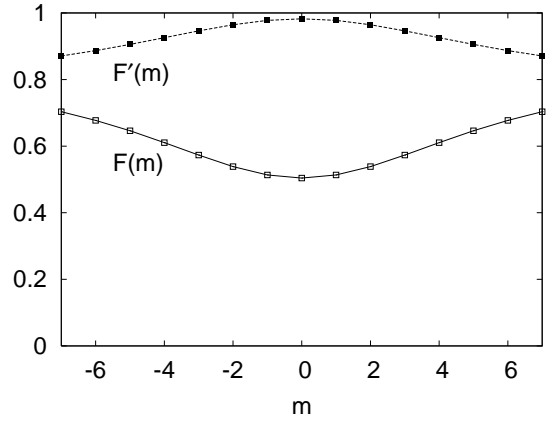}
\end{center}
\caption{\label{fig2}Fidelity $F(m)$ after measurement $\{\hat{M}_m\}$
and mean fidelity $F'(m)$ after
the Hermitian conjugate measurement $\{\hat{C}^{(m)}_\mu\}$ as functions of
the first outcome $m$,
with $s=1/2$, $j=7$, $g=0.25$, and $\theta=\pi/6$.
}
\end{figure}

\begin{figure}
\begin{center}
\includegraphics[scale=0.65]{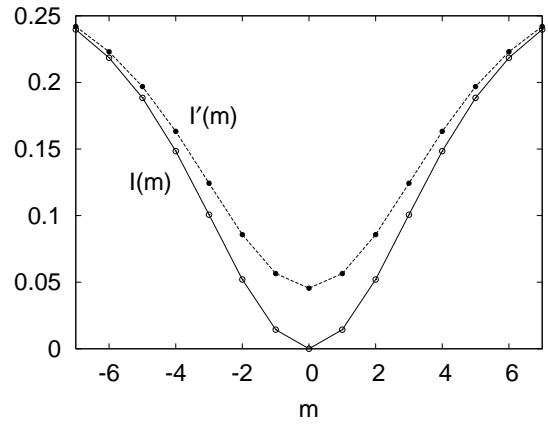}
\end{center}
\caption{\label{fig3}Information $I(m)$ after measurement $\{\hat{M}_m\}$
and mean information $I'(m)$ after
the Hermitian conjugate measurement $\{\hat{C}^{(m)}_\mu\}$ as functions of
the first outcome $m$,
with $s=1/2$, $j=7$, $g=0.25$, and $\theta=\pi/6$.}
\end{figure}

Figures \ref{fig1}, \ref{fig2}, and \ref{fig3} show $p(m)$, $F(m)$,
and $I(m)$, respectively,
as functions of $m$ for $s=1/2$, $j=7$, $g=0.25$,
and $\theta=\pi/6$, where the assumptions
in Eqs.~(\ref{eq:assump1}) and (\ref{eq:assump2}) are satisfied.
In Fig.~\ref{fig3}, $I(m)$ deviates from Eq.~(\ref{eq:exampleim})
for large $|m|$, since higher-order terms in $g$ are not
negligible there.
Note that $\hat{\epsilon}_m$ for $|m|\simeq j$ is not necessarily small
even if Eq.~(\ref{eq:assump1}) is assumed,
though the probability for such $m$ is very small,
as shown in Fig.~\ref{fig1}.
The mean fidelity and information gain are
$F=0.535$ and $I=0.045$, respectively.
In this example,
Eq.~(\ref{eq:assumpfm}) is satisfied when $-5\le m\le 5$.

To recover the original state $\hat{\rho}(a)$,
we next perform a Hermitian conjugate measurement
on the state $\hat{\rho}(m,a)$
after measurement $\{\hat{M}_m\}$.
It is chosen independently of $m$ as
\begin{equation}
\hat{C}^{(m)}_\mu=\hat{T}_\mu(\pi-\theta)=\sum_\sigma
  a_{\mu\sigma}^{(j)}(\pi-\theta)\,
\ket{\sigma}\bra{\sigma},
\end{equation}
which can be achieved in the same way as
the measurement $\{\hat{M}_m\}$,
by replacing the initial probe state $\ket{\theta,\pi/2}$ with
$\ket{\pi-\theta,\pi/2}$.
The preferred outcome $\mu_0$ is equal to $m$, because
\begin{equation}
 \hat{T}_{m}(\pi-\theta)= (-1)^{j+m}
   \,\hat{T}_{m}^\dagger(\theta).
\end{equation}
Note that this measurement $\{\hat{C}^{(m)}_\mu\}$ can also
be regarded as a reversing measurement
with the preferred outcome $\nu_0=-m$ if $s=1/2$~\cite{TerUed05}, since
\begin{equation}
 \hat{T}_{-m}(\pi-\theta)\propto
   \, \hat{T}_{m}^{-1}(\theta);
\end{equation}
this relation holds only approximately if $s>1/2$.
In fact, an approximate recovery with additional information gain
was first reported~\cite{TerUed05} regarding the reversing measurement
without identifying the origin of the information gain.
The origin is now clarified in terms of
the Hermitian conjugate measurement.
If the initial probe state for $\hat{M}_m$ is
the more general $\ket{\theta,\phi}$~\cite{TerUed05},
that for the Hermitian conjugate measurement is
$\ket{\pi-\theta,\phi}$ with $\mu_0=m$ or
$\ket{\pi-\theta,\phi+\pi}$ with $\mu_0=-m$,
while that for the reversing measurement of $s=1/2$ is
$\ket{\pi-\theta,\pi-\phi}$ with $\nu_0=-m$ or
$\ket{\pi-\theta,-\phi}$ with $\nu_0=m$.

\begin{figure}
\begin{center}
\includegraphics[scale=0.65]{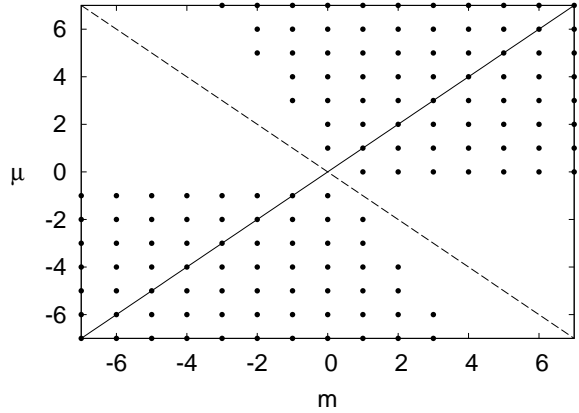}
\end{center}
\caption{\label{fig4}Sets of outcomes $(m,\mu)$
for which $F(m,\mu)>F(m)$ and $I(m,\mu)>I(m)$,
with $s=1/2$, $j=7$, $g=0.25$, and $\theta=\pi/6$.
The solid line ($\mu=m$) denotes the Hermitian conjugate measurement with
the preferred outcome, while the dashed line ($\mu=-m$)
corresponds to the reversing measurement
with the preferred outcome, $F(m,-m)=1$ and $I(m,-m)=0$.}
\end{figure}

If the Hermitian conjugate measurement $\{\hat{C}^{(m)}_\mu\}$
yields an outcome $\mu$ ($=-j,-j+1,\ldots,j-1,j$),
the fidelity and information gain become
\begin{align}
 F(m,\mu) &\simeq 1-\frac{1}{3}g^2s(2s+1)(\mu+m)^2\sin^2\theta, \\
 I(m,\mu) &\simeq \frac{4}{3}g^2s(\mu+m)^2\sin^2\theta.
\end{align}
Figure \ref{fig4} plots the sets of outcomes $(m,\mu)$
for which $F(m,\mu)>F(m)$ and $I(m,\mu)>I(m)$
with $s=1/2$, $j=7$, $g=0.25$, and $\theta=\pi/6$.
The conditional probability for
the preferred outcome $\mu_0=m$ in Eq.~(\ref{eq:condprob})
is shown in Fig.~\ref{fig1}.
Taking the average over outcome $\mu$,
we obtain the mean fidelity and mean information defined in
Eqs.~(\ref{eq:deffmp}) and (\ref{eq:defimp}), respectively, as
\begin{align}
 F'(m) &\simeq 1-\frac{1}{3}g^2s(2s+1)
          \left(m^2+\frac{j}{2}\right) \sin^2\theta, \\
 I'(m) &\simeq \frac{4}{3}g^2s
          \left(m^2+\frac{j}{2}\right) \sin^2\theta.
\end{align}
Figures \ref{fig2} and \ref{fig3} also show $F'(m)$ and $I'(m)$,
respectively, as functions of $m$.
Note that, in this example,
$F'(m)>F(m)$ and $I'(m)>I(m)$ for any value of $m$.
If the average over outcome $m$ is taken,
the total mean fidelity in Eq.~(\ref{eq:deffbarp})
and total mean information in Eq.~(\ref{eq:defibarp})
are given by
\begin{align}
F' &\simeq
 1-\frac{1}{3}g^2s(2s+1)j\sin^2\theta, \label{eq:weakfp} \\
I' &\simeq
   \frac{4}{3}g^2sj\sin^2\theta.
\end{align}
Assumption (\ref{eq:assump1}) ensures that
$F'$ is close to $1$.
Unlike Eq.~(\ref{eq:weakf}),
no term of order $g^2j^2$ appears
in the fidelity expression in Eq.~(\ref{eq:weakfp}),
because the effect of large $\hat{\Gamma}_m$ is canceled
out by the Hermitian conjugate measurement.
When $s=1/2$, $j=7$, $g=0.25$, and $\theta=\pi/6$,
$F'=0.966>F$ and $I'=0.081>I$.
Thus, the Hermitian conjugate measurement
increases both fidelity and information gain
when the particular outcomes are obtained,
as well as when averages over the outcomes are taken.

\section{\label{sec:conclude}Conclusion and Discussion}
We have discussed a probabilistic reversing operation on
a system subjected to a state change caused by a weak measurement.
The reversing operation 
can increase not only the fidelity to its original state
but also the information gain.
The essential feature of the operation is to
utilize the Hermitian conjugate of
the measurement operator, rather than its inverse.
The Hermitian conjugate operator cancels the unitary part of
the measurement operator, which does not carry information,
and enhances the information-carrying nonunitary part
because the composition of $\hat{M}_m$ and $\hat{C}^{(m)}_{\mu_0}$
results in the optimal measurement $\hat{N}_m$ being applied twice,
as shown in Eq.~(\ref{eq:conjrel}).
In contrast, the inverse operator cancels both unitary and nonunitary parts.
As an explicit example, we considered
a quantum measurement of a spin-$s$ system
using a spin-$j$ probe and
demonstrated that the reversing operation can
increase not only the fidelity and information gain
with a high probability, but also their average values.
The measurement and its reversing operation described
in Sec.~\ref{sec:example} can be implemented~\cite{TerUed05} using
an ensemble of $2s$ two-level atoms as a system
and a collection of $2j$ photons with two polarizations
(horizontal or vertical) as a probe.
The interaction in Eq.~(\ref{eq:int}) is then realized
via a Faraday
rotation~\cite{HapMat67,KuBiMa98,THTTIY99,KMJYEB99}.

The Hermitian conjugate measurement $\{\hat{C}^{(m)}_\mu\}$
is more feasible than
the reversing measurement $\{\hat{R}^{(m)}_\nu\}$.
Consider a quantum measurement in which a probe with
initial state $\ket{i}$ interacts with the system
via an interaction $\hat{U}_{\mathrm{int}}$,
and then it is measured with respect to
a certain observable.
The measurement operator for this measurement is written as
$\hat{M}_m=\bra{m}\hat{U}_{\mathrm{int}}\ket{i}$,
where $\ket{m}$ is
the final state of the probe corresponding to outcome $m$.
Since its Hermitian conjugate operator is given by
$\hat{M}_m^\dagger=\bra{i}\hat{U}_{\mathrm{int}}^\dagger\ket{m}$,
the Hermitian conjugate measurement can be performed by
a probe with initial state $\ket{m}$ together with
the time-reversed interaction $\hat{U}_{\mathrm{int}}^\dagger$.
The preferred outcome is the one
that corresponds to the probe state $\ket{i}$.
The implementation of the Hermitian conjugate measurement can
be complicated in more general situations.
Nevertheless, in photon counting~\cite{UeImOg90},
the standard photon counter implements
the annihilation operator $\hat{a}$ of the photon,
while the quantum
counter~\cite{UedKit92,UeImNa96,Bloemb59,Mandel66,UNSTN04}
implements its Hermitian conjugate operator,
i.e., the creation operator $\hat{a}^\dagger$.

Note that, while the Hermitian conjugate of an operator
always exists, unlike the inverse,
it does not always increase the fidelity and information gain.
For example, a projection operator $\hat{P}$ does not have
an inverse $\hat{P}^{-1}$,
but it does have the Hermitian conjugate $\hat{P}^\dagger=\hat{P}$.
However, when the measurement operator $\hat{M}_m$
is a projection operator, the Hermitian conjugate measurement
leaves the fidelity and information gain unchanged.
Moreover, in the case of an optimal measurement $\{\hat{N}_m\}$,
its Hermitian conjugate measurement increases the information gain
but decreases the fidelity.
Thus, our approximate recovery with additional information gain
relies on assumptions in Eqs.~(\ref{eq:weakn}) and (\ref{eq:assumpfm}),
which mean that the measurement provides little information
but drastically changes the state of the system
because $\hat{\epsilon}_m$ is small and $\hat{\Gamma}_m$ is large.

It might appear that our conclusion is due to
the choice of information measure in Eq.~(\ref{eq:defim}).
However, the same conclusion could be drawn from
another appropriate measure of information,
such as the measure proposed in Ref.~\cite{BuHaHo07}.
This is because Eq.~(\ref{eq:conjrel}) states that
the combined effect of operations of 
$\hat{M}_m$ and $\hat{C}^{(m)}_{\mu_0}$
amounts to applying the optimal measurement $\hat{N}_m$ twice.
If we perform a measurement twice and obtain the same outcome,
our knowledge about the state of the system becomes more
accurate than for a single measurement outcome.

In quantum cryptography~\cite{BenBra84,Ekert91,Bennet92,BeBrMe92},
our scheme could benefit eavesdroppers.
If the available interactions are limited,
the information obtained by eavesdropping would be lowered
with respect to the disturbance of the state transferred
between the sender and the receiver.
However, the Hermitian conjugate measurement could make
eavesdropping more efficient,
since it approximately recovers the state
with additional information gain.
On the other hand,
in quantum error-correction~\cite{Shor95,Steane96,LMPZ96},
the Hermitian conjugate measurement scheme has less advantage than
the reversing measurement scheme~\cite{KoaUed99}, since
no information gain is required,
and the emphasis is on perfect state recovery.

\section*{Acknowledgments}
This research was supported by a Grant-in-Aid
for Scientific Research (Grant No.~20740230) from
the Ministry of Education, Culture, Sports,
Science and Technology of Japan.

\appendix
\section*{Appendix}

\section{\label{sec:vari}Calculation of Variances}
We here prove Eqs.~(\ref{eq:varii}) and (\ref{eq:varif}).
The variances are defined by
\begin{align}
 V_I(\hat{S}_z) &= \overline{\expv{\hat{S}_z}^2}-
     \left(\,\overline{\expv{\hat{S}_z}}\,\right)^2, \label{eq:defvarii}\\
 V_F(\hat{S}_z) &= \overline{\expv{\hat{S}_z^2}}-
     \overline{\expv{\hat{S}_z}^2}, \label{eq:defvarif}
\end{align}
where the expectation values are given
from Eqs.~(\ref{eq:overline}) and (\ref{eq:pure}) by
\begin{align}
   \overline{\expv{\hat{S}_z}} &=
     \frac{1}{N}\sum_a\sum_\sigma |c_\sigma(a)|^2 \sigma, \\
   \overline{\expv{\hat{S}_z^2}} &= 
     \frac{1}{N}\sum_a\sum_\sigma |c_\sigma(a)|^2 \sigma^2, \\
   \overline{\expv{\hat{S}_z}^2} &= 
     \frac{1}{N}\sum_a\sum_{\sigma,\sigma'} 
        |c_\sigma(a)|^2|c_{\sigma'}(a)|^2  \sigma \sigma'.
\end{align}
Since index $a$ runs over all pure states,
there is no preferred $\sigma$.
From this symmetry, we can set
\begin{equation}
   \frac{1}{N}\sum_a |c_\sigma(a)|^2 \equiv C
\end{equation}
and
\begin{equation}
 \frac{1}{N}\sum_a
        |c_\sigma(a)|^2|c_{\sigma'}(a)|^2 \equiv
   \begin{cases}
      D & \mbox{(if $\sigma=\sigma'$)}; \\
      E & \mbox{(if $\sigma\neq\sigma'$)},
   \end{cases}
\end{equation}
where $C$, $D$, and $E$ are constants that do not
depend on $\sigma$ and $\sigma'$.
Using these constants with
the summations $\sum_\sigma \sigma =0$ and
\begin{equation}
  \sum_\sigma  \sigma^2 =-\sum_{\sigma\neq\sigma'}\sigma\sigma'
   =\frac{1}{3}s(s+1)(2s+1),
\end{equation}
it can be shown that
\begin{align}
   \overline{\expv{\hat{S}_z}} &= 0, \label{eq:expv1} \\
   \overline{\expv{\hat{S}_z^2}} &= 
     \frac{1}{3}s(s+1)(2s+1)C,  \\
   \overline{\expv{\hat{S}_z}^2} &=
     \frac{1}{3}s(s+1)(2s+1)(D-E). \label{eq:expv3}
\end{align}

To calculate $C$, $D$, and $E$,
we introduce a parametrization of coefficients $\{c_\sigma(a)\}$.
Let $\alpha_\sigma(a)$ and $\beta_\sigma(a)$ be
the real and imaginary parts of $c_\sigma(a)$, respectively.
The normalization condition then becomes
\begin{equation}
  \sum_\sigma |c_\sigma(a)|^2=
   \sum_\sigma \left[\alpha_\sigma(a)^2+\beta_\sigma(a)^2\right]=1,
\end{equation}
which is the condition for a point to be on
the unit sphere in $2(2s+1)$ dimensions.
Therefore, we parametrize
$\alpha_\sigma(a)$ and $\beta_\sigma(a)$
using hyperspherical coordinates as
\begin{align}
   \alpha_s(a) &= \sin\theta_{4s}\sin\theta_{4s-1}\cdots
                   \sin\theta_3\sin\theta_2\sin\theta_1\cos\phi, \notag \\
   \beta_s(a) &= \sin\theta_{4s}\sin\theta_{4s-1}\cdots
                   \sin\theta_3\sin\theta_2\sin\theta_1\sin\phi, \notag  \\
   \alpha_{s-1}(a) &= \sin\theta_{4s}\sin\theta_{4s-1}\cdots
                   \sin\theta_3\sin\theta_2\cos\theta_1, \notag  \\
   \beta_{s-1}(a) &= \sin\theta_{4s}\sin\theta_{4s-1}\cdots
                   \sin\theta_3\cos\theta_2,  \\
             &\vdots   \notag \\
   \alpha_{-s}(a) &= \sin\theta_{4s}\cos\theta_{4s-1}, \notag  \\
   \beta_{-s}(a) &= \cos\theta_{4s}, \notag
\end{align}
with $0\le \phi < 2\pi$ and
$0\le \theta_p \le \pi$ ($p=1,2,\ldots,4s$).
By replacing the summation over $a$ with an integral,
\begin{equation}
   \frac{1}{N}\sum_a   \quad\longrightarrow\quad
     \frac{(2s)!}{2\pi^{2s+1}}\int^{2\pi}_0 d\phi\,
     \prod^{4s}_{p=1}\int^\pi_0 d\theta_p\sin^p \theta_p,
\end{equation}
and setting $\sigma=s$ and $\sigma'=-s$, we find that
\begin{align}
  C &= \frac{1}{N}\sum_a |c_s(a)|^2 
    = \frac{(2s)!}{\pi^{2s}} \prod^{4s}_{p=1}\int^\pi_0 d\theta_p
                 \sin^{p+2} \theta_p, \\
  D &= \frac{1}{N}\sum_a |c_s(a)|^4 
    = \frac{(2s)!}{\pi^{2s}} \prod^{4s}_{p=1}\int^\pi_0 d\theta_p
                 \sin^{p+4} \theta_p, \\
   E &= \frac{1}{N}\sum_a |c_s(a)|^2 |c_{-s}(a)|^2 \notag \\
    &= C-\frac{(2s)!}{\pi^{2s}}
\prod^{4s}_{p=4s-1}\int^\pi_0 d\theta_p
                 \sin^{p+4} \theta_p \times
   \prod^{4s-2}_{p=1}\int^\pi_0 d\theta_p
                 \sin^{p+2} \theta_p.
\end{align}
Using the integral formula
\begin{equation}
  \int^\pi_0 d\theta \,\sin^n \theta =\sqrt{\pi}\,
  \frac{\Gamma\left(\frac{n+1}{2}\right)}
      {\Gamma\left(\frac{n+2}{2}\right)}
\end{equation}
for $n>-1$ with the Gamma function $\Gamma(n)$,
the constants are calculated to be
\begin{equation}
  C = \frac{1}{2s+1},  \quad
  D = \frac{1}{(s+1)(2s+1)},  \quad
  E = \frac{1}{2(s+1)(2s+1)}.
\end{equation}
Substituting these results
into Eqs.~(\ref{eq:expv1})--(\ref{eq:expv3}),
we finally obtain
\begin{equation}
   \overline{\expv{\hat{S}_z}} = 0, \qquad
   \overline{\expv{\hat{S}_z^2}} = \frac{1}{3}s(s+1),\qquad
   \overline{\expv{\hat{S}_z}^2} =  \frac{1}{6}s,
\end{equation}
which prove Eqs.~(\ref{eq:varii}) and (\ref{eq:varif})
through definitions (\ref{eq:defvarii}) and (\ref{eq:defvarif}).


\end{document}